\documentclass[10pt,letterpaper]{article}

\usepackage{cogsci}

\cogscifinalcopy 

\usepackage{pslatex}
\usepackage{apacite}
\usepackage{float} 

\usepackage{subcaption}
\usepackage{graphicx}
\graphicspath{{figures}}
\usepackage{makecell}
\usepackage{bm}
\usepackage{multirow}
\usepackage{footmisc}
\usepackage[table]{xcolor}
\usepackage{hhline}
\usepackage{wrapfig}

\newlength{\Oldarrayrulewidth}
\newcommand{\Cline}[2]{%
  \noalign{\global\setlength{\Oldarrayrulewidth}{\arrayrulewidth}}%
  \noalign{\global\setlength{\arrayrulewidth}{#1}}\cline{#2}%
  \noalign{\global\setlength{\arrayrulewidth}{\Oldarrayrulewidth}}}

\title{Preparing Unprepared Students  For Future Learning}
 
\author{{ \large \bf Mark Abdelshiheed, Mehak Maniktala, Song Ju, Ayush Jain, Tiffany Barnes, and Min Chi} \\
  Department of Computer Science\\
  North Carolina State University\\
  Raleigh, NC 27695 \\
  \{{mnabdels,\, mmanikt,\, sju2,\, ajain37,\, tmbarnes,\, mchi\}}@ncsu.edu}

\begin{document}

\maketitle

\begin{abstract}

 Based on \emph{strategy-awareness} (knowing \emph{which} problem-solving strategy to use) and \emph{time-awareness} (knowing \emph{when} to use it), students are categorized into \emph{Rote} (neither type of awareness), \emph{Dabbler} (strategy-aware only) or \emph{Selective} (both types of  awareness). It was shown that Selective is often significantly more prepared for future learning than Rote and Dabbler \cite{abdelshiheed2020metacognition}. In this work,  we explore the impact of \emph{explicit strategy instruction} on Rote and Dabbler students across two domains: logic and probability.  During the logic instruction, our logic tutor handles both Forward-Chaining (FC) and Backward-Chaining (BC) strategies, with FC being the default;  the Experimental condition is taught \emph{how to use BC via worked examples} and \emph{when to use it via prompts}. Six weeks later, all students are trained on a probability tutor that supports BC only.  Our results show that Experimental significantly outperforms Control in both domains, and Experimental Rote catches up with Selective.


\end{abstract}
\section{Introduction}

It is commonly observed that certain learners are better prepared for future learning in that they are less sensitive to learning environments and can always learn; while others are more sensitive to variations in learning environments and may fail to learn \cite{song2021AIED,markelATI, zhou2017impact,chi2010backward2metacogStrategyINSTRUCTION, cronbach1977aptitudes}. It is not fully understood why such differences exist and one of many hypotheses is that certain learners lack specific skills such as general problem-solving strategies or metacognitive skills \cite{abdelshiheed2020metacognition,pintrich2002roleSWITCH, hartman2001developingSWITCH}. 

In this work, we focus on two types of metacognitive skills related to problem-solving strategies: \emph{strategy-awareness} and \emph{time-awareness}, that is,  knowing \emph{which strategy} to use and \emph{when} to use it.   Based on these two types, we classify students into three groups: ``\textbf{\emph{Rote}}'' referring to those who never switch problem-solving strategies and simply follow the default strategy, ``\textbf{\emph{Dabbler}}'' for those who seemingly know different problem-solving strategies but not proficient enough to know \emph{when} to use each, and finally, ``\textbf{\emph{Selective}}'' refers to students who know both \emph{which} strategy to use and \emph{when} to use it. In our prior work, we analyzed data from $495$ college students who studied logic and then probability using two intelligent tutoring systems (ITSs) and we found that Selective outperformed both Rote and Dabbler across both tutors \cite{abdelshiheed2020metacognition}. 
These findings were consistent with prior research, in that skilled problem solvers tend to understand how, when, and why to use each strategy \cite{de2018longSWITCH-INSTRUCTION, pintrich2002roleSWITCH, alexander1998perspectiveSWITCH}. 
In this work, we explore whether  \emph{explicit instruction on \textbf{how} and \textbf{when} to use a problem-solving strategy} in one domain can make Rote and Dabbler catch up with Selective using two intelligent tutoring systems (ITSs) across two deductive domains: logic and probability.

In deductive task domains,  solving a problem often requires deriving an argument or proof consisting of one or more inferential steps, and each step is the result of applying a domain principle, operator, or rule. Two common problem-solving strategies in deductive domains are forward chaining (FC) and backward chaining (BC) \cite{russell2010artificial}. Prior work has shown that students often use a mixture of FC and BC during their problem-solving \cite{priest1992newINSTRUCTION,simon1978individualINSTRUCTION}. In this work, our logic tutor handles both FC and BC strategies but defaults to FC, while our probability tutor supports BC only. During the logic instruction, the Experimental condition is explicitly taught BC strategy through \emph{worked examples} (strategy-awareness) and the tutor would \emph{prompt} them to switch to it when it is proper to do so (time-awareness),  while the Control condition is trained on the same tutor without any explicit strategy instruction. After six weeks, both conditions are trained on a probability tutor. Our results show that the Experimental condition significantly outperforms Control in both domains. More importantly,  the Rote students benefit the most from our intervention, as they significantly outperform their peers on the logic tutor and continue to perform well in the probability tutor. By analyzing their strategy switch behavior, we find that they indeed catch up with their Selective peers, who know in advance which strategy to use when.

\section{Related Work}

Generally speaking, metacognition indicates one's ability to regulate, understand and monitor their cognitive skills \cite{chambres2002metacognitionStrategySelection,roberts1993metacogDefinitionStrategySelection2}. Many studies have investigated the impact of teaching metacognitive skills on student learning \cite{zepeda2015INSTRUCTION, chi2010backward2metacogStrategyINSTRUCTION} and explored different ways of teaching them effectively \cite{de2018longSWITCH-INSTRUCTION,cardelle1992effectsMETACOGNITIVE}. Based on \citeA{winne2014switchMetacognitiveSWITCH},  mastering strategy selection alone is a cognitive skill, but when incorporated with awareness about when such strategy should be used, it can be considered as a metacognitive skill. Thus in this work, we consider two types of awareness: \emph{strategy-awareness} and \emph{time-awareness}, which are respectively,  which strategy to use and when to use it \cite{de2018longSWITCH-INSTRUCTION}. 

\subsection{Strategy Instruction}

Researchers have explored a variety of instruction types for teaching strategies \cite{de2018longSWITCH-INSTRUCTION,likourezos2017instruction,zepeda2015INSTRUCTION,sporer2009improvingINSTRUCTION}. For instance, \citeA{sporer2009improvingINSTRUCTION} analyzed the role of explicit instruction of multiple reading strategies on the comprehension of third- to sixth-graders. They found that students who were instructed explicitly outperformed their peers, who were taught traditionally by the instructors' text interactions, on a transfer posttest task and follow-up test.  

In this work, we focus on two problem-solving strategies: FC and BC. In the former, students progress from givens to the goal, while in the latter, they reduce the goal to givens. Prior research has highlighted the impact of teaching \textit{FC} \cite{shrestha2013usingINSTRUCTION}, \textit{BC} \cite{chi2010backward2metacogStrategyINSTRUCTION}, or both strategies \cite{priest1992newINSTRUCTION,larkin1980expertINSTRUCTION,simon1978individualINSTRUCTION}. For example, \citeA{priest1992newINSTRUCTION} compared how experts and novices solve physics problems. They found that both groups used a similar mixture of forward and backward chaining. However, \emph{only} the experts knew when and why to use each strategy, as they significantly produced more complete plans and stages compared to their novice peers.  As far as we know, prior work taught FC or BC by asking students to solve problems with/without feedback or support. In this work, we use \emph{worked examples} to teach students BC in the logic tutor, where students are allowed to choose either FC or BC to solve each problem.

\subsection{The Worked-Example Effect}

A worked example (WE) is a provided step-by-step solution that solves a problem or completes a task. In cognitive load theory, the \emph{worked-example effect} refers to the observed learning outcome from teaching with WEs \cite{sweller1985WE}. Substantial work has leveraged WEs to enhance students' problem-solving skills \cite{renkl2010learningINSTRUCTION,van2004processINSTRUCTION,paas1994variabilityINSTRUCTION}, prepare them for direct instruction \cite{likourezos2017instruction}, seal knowledge-gap experience \cite{glogger2015INSTRUCTION}, reduce cognitive load \cite{gerjets2004designingINSTRUCTION}, and deepen the conceptual understanding \cite{schwonke2009WE}. 

\citeA{renkl2010learningINSTRUCTION} addressed when to present WEs to students. They found that WEs contribute positively to the learning performance when offered in the \emph{early} stages of skill acquisition. However, students will likely stop paying attention to them in later stages. The authors argued that WEs should elicit some steps from students, and possess self-explanatory prompts to strengthen their impacts. Furthermore, \citeA{schwonke2009WE} investigated the impact of faded worked examples (FWEs), which is a mixture of solution steps and incomplete steps to be solved by students. They found that students who were given FWEs were able to learn more efficiently and show deeper conceptual knowledge of geometry principles, compared to their peers who had supportive tutoring in the form of corrective feedback and self-explanation prompts.

In short, given that WEs have shown great promise of improving student content learning, in this work, we use WEs in the early stages of the logic tutor to teach the \emph{BC} strategy.

\subsection{The Importance of ``When to use Which Strategy''}

Prior work suggests that students often lack metacognitive knowledge because they do not understand \emph{when or why} to use a specific strategy \cite{de2018longSWITCH-INSTRUCTION, pintrich2002roleSWITCH, hartman2001developingSWITCH}, and thus in order to raise their metacognitive awareness, they must master such  skills \cite{winne2014switchMetacognitiveSWITCH, alexander1998perspectiveSWITCH}. For instance, \citeA{de2018longSWITCH-INSTRUCTION} investigated the long-term effects of metacognitive strategy instruction on students' academic performance. They found that students who were given interventions that include when, why, how, and which strategy to use, showed superior planning skills. Moreover, their learning performance was the highest, as they did not only outperform their peers on a posttest task, but also a far follow-up test. \citeauthor{de2018longSWITCH-INSTRUCTION} argued that in multi-strategy domains, it is insufficient to only learn \textit{what} each strategy is. Rather, it is equally important to learn \textit{when} to use each.

On the other hand, compared with a large amount of prior work on explicit strategy instruction, little research has been done on how to teach students to become time-aware. Some related work has shed light on this subject. For example, it has been shown that young adults ($<30$ years) are more flexible in and capable of, revising their initial strategy choice for a better strategy \cite{taillan2015relationshipsSWITCH,ardiale2012withinSWITCH}. More specifically, \citeA{taillan2015relationshipsSWITCH} compared the strategy switch behavior between young and older adults in a computational task and found that the former were more receptive to trying new strategies. Moreover, \citeauthor{taillan2015relationshipsSWITCH} argued that the likelihood of switching a strategy is much higher when all strategies have similar difficulty.  Inspired by these findings,  we use \emph{prompts} to recommend students to switch their problem-solving strategy when it is proper to do so.

\section{Methods}
\subsection{Two Tutors and Our  Interventions}
\subsubsection{Logic Tutor and Instructional Interventions}

\begin{figure}[ht!]
     \centering
     \begin{subfigure}[t]{0.23\textwidth}
         \centering
         \includegraphics[width=\textwidth,height =4cm]{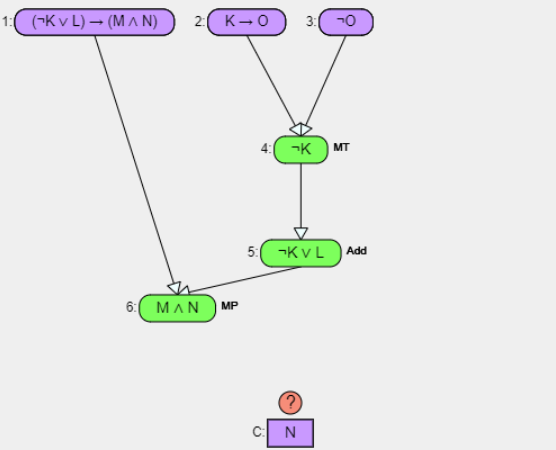}
         \caption{Forward Chaining}
         \label{fig:direct}
     \end{subfigure}
     \hfill
     \begin{subfigure}[t]{0.24\textwidth}
         \centering
         \includegraphics[width=\textwidth,height=4cm]{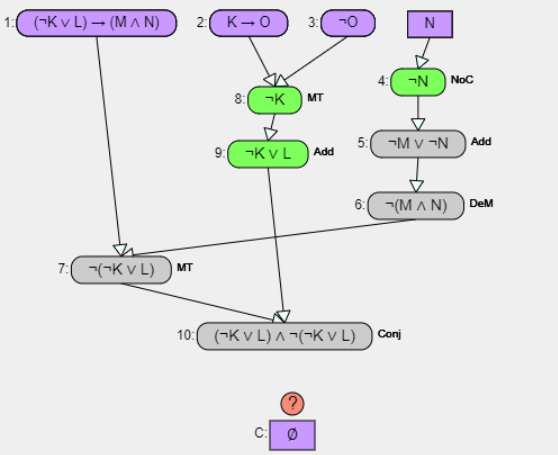}
         \caption{Backward Chaining}
         \label{fig:indirect}
     \end{subfigure}
\caption{Logic Tutor Problem-Solving Strategies}
\label{DT}
\end{figure}

\begin{figure}[ht]
\begin{center}
\includegraphics[width=0.4\textwidth]{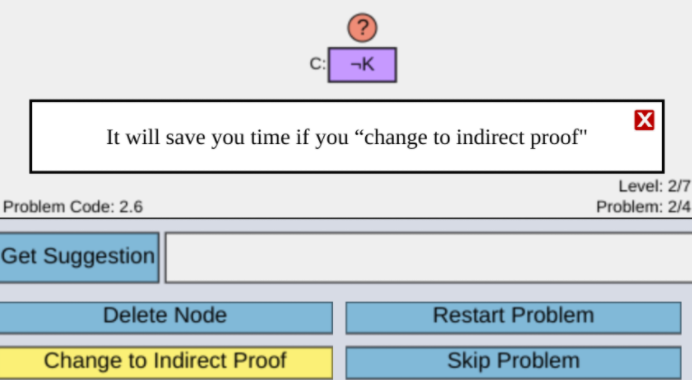}
\end{center}
\caption{Strategy Switch Prompt} 
\label{fig:prompt}
\end{figure}

Our logic tutor teaches students propositional logic proofs.  It consists of five ordered levels with an \emph{incremental degree of difficulty} and each level consists of four problems.  A student can solve any problem by either a \textbf{FC} or \textbf{BC} strategy. Figure \ref{fig:direct} shows that for \emph{FC}, one must derive the conclusion at the bottom from givens at the top;  while Figure \ref{fig:indirect} shows that for \emph{BC}, students need to derive a contradiction from givens and the \emph{negation} of the conclusion. Problems are presented by \emph{default} in FC, but students can switch to BC by clicking the yellow button in Figure \ref{fig:prompt}.

\begin{figure}[ht!]
\begin{center}
\includegraphics[width=8cm]{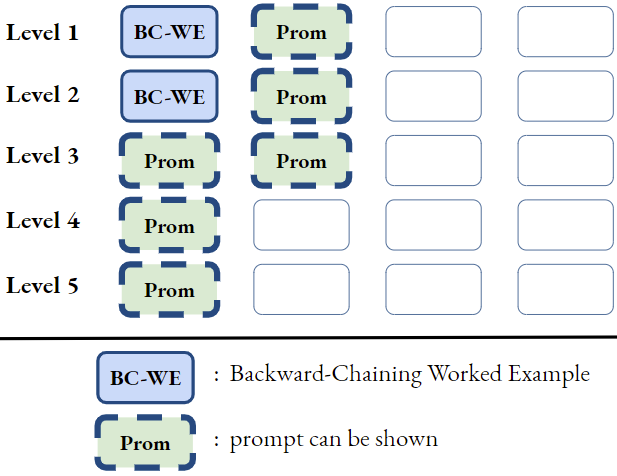}
\end{center}
\caption{Training on the Modified Logic Tutor} 
\label{fig:procedure}
\end{figure}

For the purpose of this work,  we modified our logic tutor by implementing two WEs on explicit BC strategy instruction and by offering \emph{advisory} prompts (the text in the black box in Figure \ref{fig:prompt}) to switch to the BC strategy when it is proper to do so. Figure \ref{fig:procedure} shows that the two BC WEs are presented as the first problem at Levels 1 and 2; all problems in green are `proper' problems to be solved by BC. We used the strategy switch behavior from our prior study \cite{abdelshiheed2020metacognition} to guide us \emph{which `proper'} problems to display the prompts in, by picking the most frequently switched problems, and \emph{when} to display them, by learning a probability distribution of the duration lengths that students take before switching. More specifically,  $55\%$ of the time the tutor would wait for $1.5$ minutes, $35\%$ for $3$ mins, and only $10\%$ for $6$ mins. For the remaining problems (colored in white in Figure \ref{fig:procedure}), the tutor behaves the same as the original tutor.

Our goal in this work is to investigate whether explicit BC strategy instruction using WEs combined with prompts would make Rote and Dabbler catch up with  Selective. We expect that the former two groups would benefit from our instructional interventions which are designed to scaffold the metacognitive skills that they lack. On the other hand, for the Selective group, we expect that providing them with additional scaffolding may interfere with their existing skills. Therefore, \emph{only} the Experimental Rote and Dabbler groups will get the treatment shown in Figure \ref{fig:procedure}, while the Control Rote and Dabbler groups and the Selective group will get no treatment.

\subsubsection{Probability tutor:} 
\begin{figure}[ht!]
\begin{center}
\includegraphics[width=8cm]{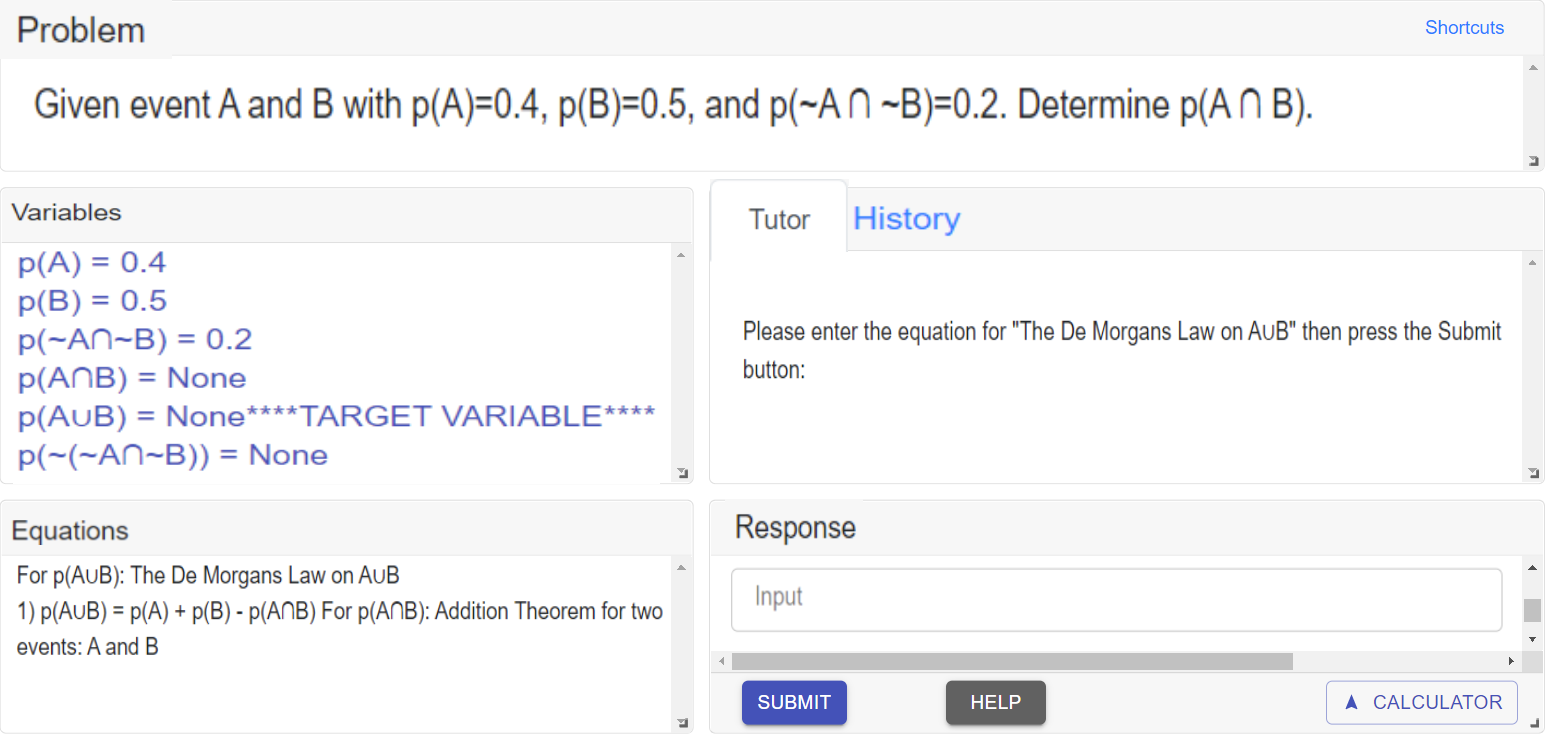}
\end{center}
\caption{Probability Tutor Interface} 
\label{fig:pyr}
\end{figure}

The interface of the tutor can be shown in Figure \ref{fig:pyr}. It is a web- and text-based tutor that teaches students how to solve mathematical probability problems using 10 major principles, such as the Complement Theorem and Bayes' Rule. It consists of $12$ problems.  Each problem can \emph{only} be solved by BC in that it requires students to derive an answer by \emph{writing and solving equations} until the target is completely reduced to the givens.

\begingroup
\renewcommand{\arraystretch}{1.7}
\begin{table}[ht!]
\footnotesize
\begin{center} 
\caption{Overview of the Study Procedure} 
\label{fullProcedure} 
\begin{tabular}{l|cc}\hline
  \multirow{7}{*}{Logic} & \multicolumn{2}{c}{Pretest (2 problems)} \\ 
 \Xcline{2-3}{0.8pt}
 
   & \multicolumn{2}{!{\vrule width 0.5pt}c !{\vrule width 0.5pt}}{\textbf{Training (20 problems)}:} \\
     & \multicolumn{2}{!{\vrule width 0.5pt}c !{\vrule width 0.5pt}}{\cellcolor{gray!60} Experimental $\Longrightarrow$ Intervention} \\
     & \multicolumn{2}{!{\vrule width 0.5pt}c !{\vrule width 0.5pt}}{\cellcolor{gray!60} Control $\Longrightarrow$ Original} \\
     & \multicolumn{2}{!{\vrule width 0.5pt}c !{\vrule width 0.5pt}}{\cellcolor{gray!60} Selective $\Longrightarrow$ Original} \\
  \Cline{0.8pt}{2-3}
  & \multicolumn{2}{c}{Posttest (6 problems, including 2 isomorphic)} \\ \cline{1-3}
  \multicolumn{3}{c}{\textbf{Six weeks later}} \\ \cline{1-3}

  \multirow{4}{*}{Prob.} & \multicolumn{2}{c}{Textbook} \\ \cline{2-3}
  & \multicolumn{2}{c}{Pretest (14 problems)} \\ \cline{2-3}
  & \multicolumn{2}{c}{Training (12 problems)} \\ \cline{2-3}
  & \multicolumn{2}{c}{Posttest (20 problems, including 14 isomorphic)} \\ \cline{1-3}

\end{tabular}
\end{center} 
\end{table}
\endgroup

\subsection{Participants}
Our participants are Computer Science undergraduate students at North Carolina State University. The tutors are assigned as a regular homework assignment and their completion is required for full credit.  Students are told that the assignment would be graded based on their demonstrated effort, not performance. A total of $128$ students completed both tutors.
 
One of the main challenges in this work is how to determine which of the three groups a student belongs to before their training on the logic tutor starts. Previously, the three metacognitive groups were determined based on a post-hoc analysis after students completed the entire training \cite{abdelshiheed2020metacognition}. 
For the purpose of this work, we trained a random forest classifier (RFC) using prior data for the task of the early prediction of the metacognitive group. More specifically, based on \emph{how} students solve the logic pretest problems, our RFC will assign them to one of the three groups before training on the logic tutor. As a result, we had $60$ Rote, $42$ Dabbler and $26$ Selective students. 

Then the Rote and Dabbler students are randomly assigned to two conditions\footnote {The difference in size is due to the fact that we gave priority to have a sufficient number of Experimental students to perform a meaningful analysis of our intervention.}:  $N=61$ for Experimental ($35$ Rote$_{Exp}$ + $26$ Dabbler$_{Exp}$) and $N=41$  for Control ($25$ Rote$_{Ctrl}$ + $16$ Dabbler$_{Ctrl}$). We found no significant difference in the distribution of Rote and Dabbler across the two conditions: $\chi^2 (1,\, N=102) = 0.13, \, \mathit{p}=.72$.  Finally, for all students who are classified as  Selective, no intervention is provided and they use the original logic tutor and are referred to as the \emph{Selective} group later on. The accuracy of our RFC is further confirmed by the students' behaviors using the Control and the Selective students since they received no intervention. Our results show that the RFC achieved $95.5\%$ accuracy, $94.9\%$ macro-recall, and $96.1\%$ macro-precision which is comparable to its performance on the training data: $95.9\%$ accuracy, $95.1\%$ macro-recall, and $96.2\%$ macro-precision.

\subsection{Procedure}\label{procedureSection}

Table \ref{fullProcedure} summarizes our procedure. During the logic instruction, all students go through the same standard pretest-training-posttest procedure. The first two posttest problems are isomorphic to the two pretest problems. The \emph{only} difference occurs during the training on the logic tutor in that the Experimental condition (Rote$_{Exp}$ and Dabbler$_{Exp}$) is given WEs with prompts (highlighted in Figure \ref{fig:procedure}), while the Control condition (Rote$_{Ctrl}$ and Dabbler$_{Ctrl}$) and the Selective group receive no such intervention.

Six weeks later, students would be all trained on the same probability tutor following the same standard procedure:  textbook, pretest, training on ITS, and posttest. In the textbook, students study the domain principles; In pre- and post-test, students solve $14$ and $20$  open-ended problems that require them to derive an answer by writing and solving one or more equations. 14 of the posttest problems are isomorphic to the pretest problems. For the training section, all students would go over the same problems in the same order. Note that for both tutors,  the posttest is \emph{much harder} than the pretest.

\subsection{Grading criteria} In logic, a problem score is based on students' time, accuracy, and solution length. Students' \emph{pre-} and \emph{post-test} scores are calculated by averaging their pre- and post-test problem scores. In probability, students' answers are graded in a double-blind manner by experienced graders using a partial-credit rubric, where grades are based \emph{only} on accuracy. The \emph{pre-} and \emph{post-test} scores are the average grades in their respective sections. For comparison purposes, all test scores are normalized to the range of $[0,100]$.

\section{Results}

\begingroup
\renewcommand{\arraystretch}{1.7}
\begin{table}[b!]
\scriptsize
\begin{center} 
\caption{Comparing the Two Conditions across Tutors} 
\label{performanceSummary} 
\begin{tabular}{ccc|c} 
\Xhline{4\arrayrulewidth}

\multicolumn{4}{c}{Logic Tutor}\\
\hline
\makecell{}   & \makecell{Experimental \\ $(N=61)$} & \makecell{Control\\ $(N=41)$ \\} & \makecell[t]{1-way ANOVA \\ \,}  \\
\hline
Pre &    \makecell{62.8 (19.6)} &  \makecell{59.4 (18.4)}  & $\mathit{p} = .38$\\ 
Iso Post  & \makecell{75.9  (2.5)}   & \makecell{62.4  (4)} & $\mathit{F}(1,100) = 15.3,\, \mathit{p} < \textbf{.001},\,\eta = .13$   \\
Iso NLG  & \makecell{0.19  (.03)}   &  \makecell{0.03  (.06)} & $\mathit{F}(1,100) = 8.2, \, \mathit{p} = \textbf{.005},\,\eta = .08$\\
Post &  \makecell{77.4  (3.6)} & \makecell{65.4  (5.2)}  &$\mathit{F}(1,100) = 38.9,\,\mathit{p} < \textbf{.001},\,\eta = .28$ \\
NLG &   \makecell{0.2  (.05)} & \makecell{0.05  (.06)} &$\mathit{F}(1,100) = 10.6, \, \mathit{p} = \textbf{.002},\,\eta = .1$ \\
Time  & \makecell{4.32  (4.9)}   & \makecell{7.23  (8)} &$\mathit{F}(1,100) = 5.2, \,\mathit{p} = \textbf{.02},\,\eta = .05$\\

\hline
\multicolumn{4}{c}{Probability Tutor}\\
\hline
Pre &   \makecell{69.6  (18.6)} &  \makecell{74.9 (15.1)}  & $\mathit{p} = .13$  \\
Iso Post  & \makecell{92.9   (2.8)}   & \makecell{84.4 (4.1)} &  $\mathit{F}(1,100) = 18.5,\,\mathit{p} < \textbf{.001},\,\eta = .16$\\
Iso NLG  & \makecell{0.4 (.05)}   &  \makecell{0.11 (.08)} &$\mathit{F}(1,100) = 24.3,\,\mathit{p} < \textbf{.001},\,\eta = .2$\\
Post &  \makecell{88.9  (5)} & \makecell{72.2  (6.1)} &$\mathit{F}(1,100) = 58.2,\,\mathit{p} < \textbf{.001},\,\eta = .37$ \\
NLG &  \makecell{0.33  (.06)} & \makecell{-0.18  (.21)} &$\mathit{F}(1,100) = 51.4,\,\mathit{p} < \textbf{.001},\,\eta = .34$\\
Time & \makecell{1.86  (.5)}  & \makecell{1.83  (.49)} & $\mathit{p} = .76$ \\

\Xhline{4\arrayrulewidth}
\end{tabular} 
\end{center} 
\end{table}
\endgroup

\begingroup
\renewcommand{\arraystretch}{1.7}
\begin{table}[ht!]
\scriptsize
\begin{center} 
\caption{Comparing The Groups' Mean(SD) Scores} 
\label{groupScores} 
\begin{tabular}{lcc|cc||c} 
\Xhline{4\arrayrulewidth}

\multicolumn{6}{c}{Logic Tutor}\\
\hline
    & \multicolumn{2}{c|}{Experimental} & \multicolumn{2}{c||}{Control} &  \\
    \cline{2-3} 
    \cline{4-5}
    &  \makecell[t]{$Rote_{Exp}$\\ $(N=35)$} & \makecell[t]{$Dabbler_{Exp}$ \\ $(N=26)$} & \makecell[t]{$Rote_{Ctrl}$\\ $(N=25)$}   & \makecell[t]{$Dabbler_{Ctrl}$ \\ $(N=16)$}  & \makecell[t]{Selective \\ $(N=26)$} \\
\hline
Pre      &   61.8 (23)    & 64.2 (14) & 60.1 (20) & 58.3 (16) & 62.3 (21) \\
Iso Post        & 78.9 (1.9)  &  71.9 (1.6) &  64.2 (3.8) & 59.4 (4.2) &  73.1 (5.3)  \\
Iso NLG       &  0.25 (.04)    &  0.1 (.04) &   0.05 (.05) &  -0.02 (.06) &  0.08 (.05) \\
Post        & 80.3 (1.7)  &  73.4 (1.5) &  64.3 (3.5)  & 67.2 (2.9) & 72.3 (5.5) \\
NLG       &  0.25 (.03)    & 0.13 (.03) &  0.02 (.04) &  0.09 (.07)  & 0.11 (.06)\\
\hline
\multicolumn{6}{c}{Probability Tutor}\\
\hline

Pre      &   67 (20)   & 73.1 (16) & 73.2 (15)  & 77.7 (15) & 70.6 (19) \\
Iso Post        &  92.5 (3.4)     &  93.5 (3.3) &  82.5 (3.9)  & 87.4 (5.8) &  91.7 (6.2) \\
Iso NLG       &  0.43 (.06)    &  0.37 (.12) &  0.09 (.21) &  0.14 (.23) &  0.37 (.16) \\
Post       &  88 (3.1)   & 90.2 (3.1) & 71.3 (3.5) & 73.5 (5.5)  & 85.8 (5.7) \\
NLG       &  0.35 (.05)   &  0.3 (.08)  & -0.16 (.23) & -0.21 (.21) & 0.24 (.15)\\
\Xhline{4\arrayrulewidth}
\end{tabular} 
\end{center} 
\end{table}
\endgroup

\subsection{Learning Performance:}

\subsubsection{Experimental vs Control} Table \ref{performanceSummary} compares the two conditions across the two tutors and it shows  the mean(SD) of students' pre- and post-test scores, isomorphic posttest, training time (in hours), and the learning performance in terms of the normalized learning gain (NLG) defined as: $(NLG = \frac{post - pre}{\sqrt{100 - pre}})$, where 100 is the maximum test score. The last column in Table \ref{performanceSummary} shows the one-way ANOVA statistical comparisons between the two conditions with their corresponding effective size $\eta$. As shown in Table \ref{performanceSummary}, while  no significant difference was found between the two conditions in the logic pretest: $\mathit{F}(1,100) = 0.8,\, \mathit{p} = .38$, and probability pretest: $\mathit{F}(1,100) = 2.7,\, \mathit{p} = .11$, Experimental significantly outperformed Control in all other aspects, except for the training time on the probability tutor. 
 
Next, we will investigate whether both Rote and Dabbler benefited from our intervention by comparing them across the two conditions. The first five columns in Table \ref{groupScores} compare the two Experimental groups (Rote$_{Exp}$ and Dabbler$_{Exp}$) against the two  Control groups (Rote$_{Ctrl}$ and Dabbler$_{Ctrl}$). 
A two-way ANOVA using condition \{Experimental, Control\} and metacognitive group \{Rote, Dabbler\}  as factors shows no significant difference among the four groups in the logic pretest: $\mathit{F}(1,98) = 0.28,\, \mathit{p} = .6$.  A two-way ANCOVA using the logic pretest as a covariate, with condition and metacognitive group as factors, finds a significant interaction effect on the logic posttest: $\mathit{F}(1,97) = 17.3,\, \mathit{p} < .0001, \, \mathit{\eta} = 0.06$. There is also a main effect of condition: $\mathit{F}(1,97) = 66.7,\, \mathit{p} < .0001, \, \mathit{\eta} = 0.23$, as the two Experimental groups significantly outperform the two Control groups.  Subsequent contrast analyses show that while no significant difference is found between the two Control groups, a significant difference is found between the two Experimental groups:  Rote$_{Exp}$ $>$ Dabbler$_{Exp}$ ($\mathit{t}(59) = 2.9,\, \mathit{p} < .01,\, \mathit{d} = 4.3$). A two-way ANOVA using the same two factors on NLGs shows similar results with the only exception that Dabbler$_{Exp}$ does not outperform the two Control groups: Rote$_{Ctrl}$ and Dabbler$_{Ctrl}$. 

In probability, a two-way ANOVA using condition \{Experimental, Control\} and metacognitive group \{Rote, Dabbler\}  as factors shows no significant difference among the four groups in the probability pretest: $\mathit{F}(1,98) = 0.05,\, \mathit{p} = .82$. Additionally, a two-way ANCOVA using the probability pretest as a covariate, with condition and metacognitive group as factors, shows no interaction effect in the posttest but a significant main effect of condition: $\mathit{F}(1,97) = 81.9,\, \mathit{p} < .0001, \, \mathit{\eta} = 0.43$ in that the two Experimental groups significantly outperform the two Control groups. Subsequent contrast analyses show that no significant difference is found neither between the two Experimental groups nor between the two Control groups. The same findings are found in probability NLGs.

In summary, our results show that our explicit strategy instruction using worked example combined with timely prompts is indeed effective in that Experimental not only outperforms the Control in logic where the intervention occurs but also the former continues to outperform the latter in probability six weeks later when there is no such intervention. More specifically, our results show that Rote students benefit more from our interventions in that  $Rote_{Exp}$ has significantly higher posttest and NLG scores compared to the two control groups in logic, while $Dabbler_{Exp}$ did not learn significantly better than the two Control groups. For probability, both $Rote_{Exp}$ and $Dabbler_{Exp}$ significantly outperform the two Control groups in terms of posttest and NLG scores, and there is no significant difference between the two Experimental groups on the same measures.

\subsubsection{Comparisons with the Selective Group} 
It is essential to note that Selective is the ``desired'' group but only about 20\% of the students are ``naturally'' classified as such in this work. To save space, the last column in Table \ref{groupScores} shows the performance of the Selective group across all measures. In this section, we will explore whether our instructional intervention is indeed effective from two aspects: 1)  whether it would make the Rote and Dabbler students in the Experimental condition catch up with the Selective group; and 2) whether, without such intervention,  the  students in the Control condition would perform worse than the Selective group.  

As for the first aspect, our overall results show that the Experimental condition performs as well as or better than the Selective group in that  no significant difference is found between the two on all measures in both logic and probability. The only exception is that the former significantly outperforms the latter in the logic posttest: $\mathit{t}(36.3) = 2.1,\, \mathit{p} = .04$.  Next, we compare the two Experimental groups, Rote$_{Exp}$ and Dabbler$_{Exp}$,  against the Selective group.   Table \ref{groupScores} shows that no significant difference is found among the three groups on both logic and probability pretests. While no significant difference is found between Selective and Dabbler$_{Exp}$ in logic,   Rote$_{Exp}$ significantly outperforms Selective in the logic posttest:  $\mathit{t}(59) = 3.2,\, \mathit{p} = .002,\, \mathit{d} = 2$ and the logic NLG: $\mathit{t}(59) = 2.3,\, \mathit{p} = .02,\, \mathit{d} = 2.9$.  In probability, no significant difference is found among Rote$_{Exp}$, Dabbler$_{Exp}$, and Selective across all measures.

\begin{figure*}[ht!]
\begin{center}
\includegraphics[width=0.7\textwidth]{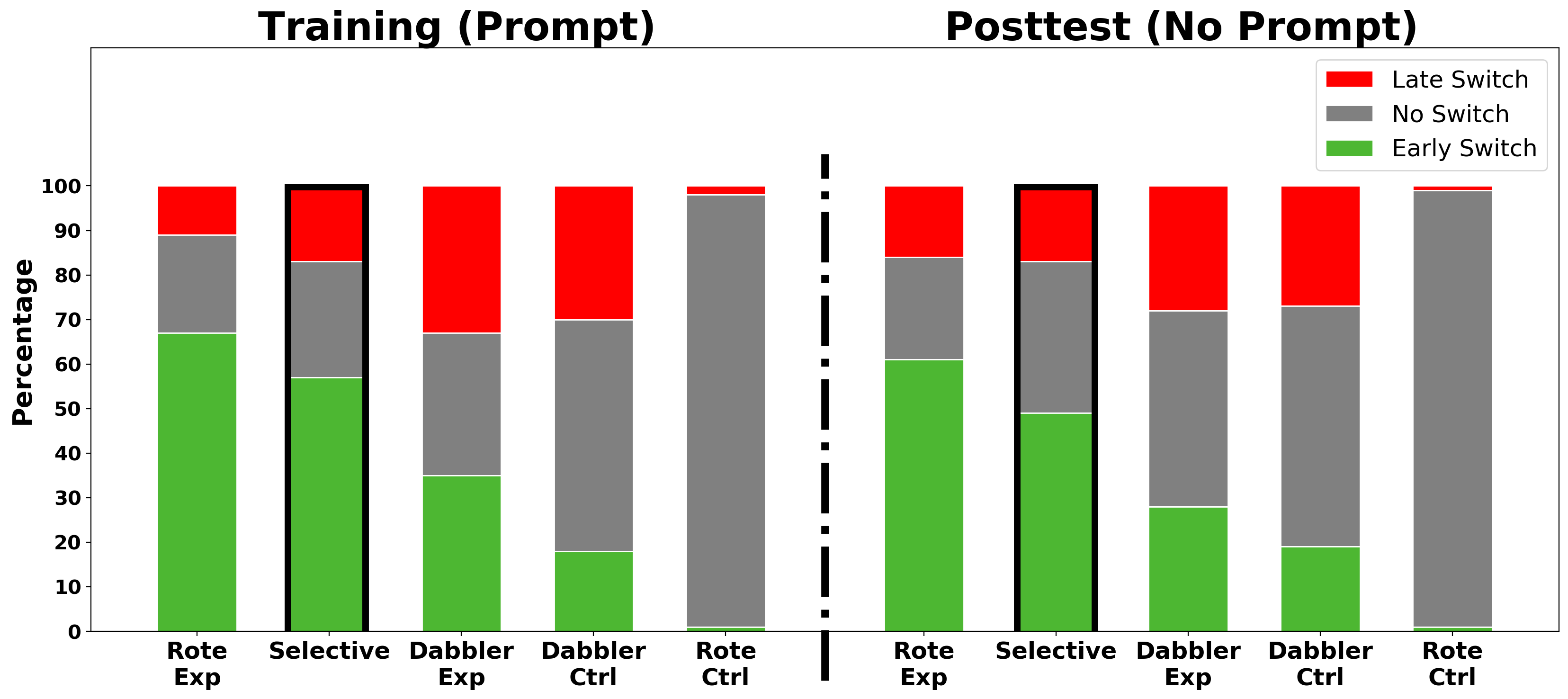}
\end{center}
\caption{Strategy Switch Behavior in The Logic Tutor} 
\label{fig:switch}
\end{figure*}

As for the second aspect, almost as expected,  the Selective group outperforms the Control condition in both domains, except for the logic NLG; In the logic posttest: $\mathit{t}(53.9) = 2.4,\, \mathit{p} = .02$, in the probability posttest: $\mathit{t}(53.4) = 4.8,\, \mathit{p} < .001$ and $\mathit{t}(63.7) = 4.4,\, \mathit{p} < .001$ for the probability NLG. Similarly, when comparing the Control groups against the Selective group,  Table \ref{groupScores} shows that: while  no significant difference is found among the three groups on both logic and probability pretests, Selective significantly outperforms the two Control groups in both logic posttest: ($\mathit{t}(49) = 3,\, \mathit{p} < .01,\, \mathit{d} = 1.7$ for Rote$_{Ctrl}$ and $\mathit{t}(40) = 3.9,\, \mathit{p} < .001,\, \mathit{d} = 1.1$ for Dabbler$_{Ctrl}$) and probability posttest: ($\mathit{t}(49) = 4.7,\, \mathit{p} < .0001 ,\, \mathit{d} = 3.1$ for Rote$_{Ctrl}$ and $\mathit{t}(40) = 3.5,\, \mathit{p} < .001 ,\, \mathit{d} = 2.2$ for Dabbler$_{Ctrl}$).

To summarize, our results show that with our instructional intervention, Experimental indeed catches up with the Selective as the former performs at least as well as Selective on both tutors.  On the other hand, without the interventions, Selective outperformed Control on the two tutors. More specifically, our results show that our interventions are especially beneficial to  Rote students in that Rote$_{Exp}$ surpassed all other groups, even Selective, in logic and continued to perform well in probability. For Dabbler, Dabbler$_{Exp}$ outperformed the two control groups in the logic posttest, but not in logic NLG; In probability, Dabbler$_{Exp}$ outperformed the two control groups.

\subsection{Strategy Switch Behavior}

Figure \ref{fig:switch} shows the strategy switch behaviors of the five groups (from FC into BC). We compared their behaviors during the logic tutor \textit{Training} where only Rote$_{Exp}$ and Dabbler$_{Exp}$ were offered prompts, and \textit{Posttest} where no student got prompts. Here, by following the same definition as described in our prior work \cite{abdelshiheed2020metacognition}, we display the percentage of \textit{No Switches} (sticking to the default strategy), \textit{Early Switches} (switching within the first $30$ actions), and \textit{Late Switches} (switching after the first $30$ actions). In Figure \ref{fig:switch}, the five groups are ordered by the percentage of \textit{Early Switches}, the most desired behavior, and  Selective is highlighted in bold as the gold standard. 

A one-way ANOVA found that the switch behaviors differed significantly among the five groups: $\mathit{F}(4,123) = 71.2,\, \mathit{p} < .0001, \, \mathit{\eta} = 0.7$ for the training and $\mathit{F}(4,123) = 62.6,\, \mathit{p} < .0001, \, \mathit{\eta} = 0.67$ for the posttest. More importantly, the behaviors of each group were very similar in the training and posttest. Subsequent contrast analyses showed that while no significant difference was observed between Rote$_{Exp}$ and Selective on their switch behaviors, both groups switched early significantly more than the other three groups: Dabbler$_{Exp}$, Dabbler$_{Ctrl}$ and Rote$_{Ctrl}$. For example, Selective switched early significantly more than Dabbler$_{Exp}$:  $\mathit{t}(50) = 5.4 ,\, \mathit{p} < .0001,\, \mathit{d} = 1.4$ for the training and $\mathit{t}(50) = 4.9,\, \mathit{p} < .0001,\, \mathit{d} = 1.3$ for the posttest.

In short, analyzing strategy switch behaviors confirms that Rote$_{Exp}$ indeed caught up with Selective, as the former showed very similar behaviors to the latter during the training when the prompts were available and more importantly, during the posttest when prompts were not present. On the other hand, much to our surprise, the strategy switch behaviors of Dabbler$_{Exp}$ stayed similar to their Control peers, Dabbler$_{Ctrl}$.

\subsection{Conclusions \& Discussions}

Ever since the theory of Preparation for Future Learning (PFL) has been proposed \cite{bransford1999transferRethinking}, substantial work has investigated how to assess PFL \cite{abdelshiheed2020metacognition,mylopoulos2016PFL,belenky2012motivationAndTransferPFL} and factors that facilitate it such as metacognitive skills \cite{zepeda2015INSTRUCTION,veenman2004relationMetacognitivePFL}, but relatively little research has been done on the design of interventions to prepare students for future learning \cite{zepeda2015INSTRUCTION}. In this work, we investigate how explicit instruction on \emph{how} and \emph{when} to use a problem-solving strategy would impact student learning across two  domains: logic  and  probability, and more importantly, whether such instruction would indeed eliminate the gap among different learners. Overall, our results show that teaching students the BC strategy using worked examples with prompts can significantly improve their learning not only in the domain they were taught but also in a new domain six weeks later. Specifically, such instructions seem to be more beneficial to students who were neither strategy-aware nor time-aware. Such findings suggest that explicit instructions on \emph{how} and \emph{when} can prepare students, especially those with little metacognitive skills, for future learning.  Despite these findings, it is important to note that there is at least one caveat in our analyses: our probability tutor only supports the BC strategy. A more convincing testbed would be to use any ITS that supports different types of strategies so we can investigate whether Rote$_{Exp}$ would continue to switch.

\section{Acknowledgments}
This research was supported by the NSF Grants: Generalizing Data-Driven Technologies to Improve Individualized STEM Instruction by Intelligent Tutors (2013502), Integrated Data-driven Technologies for Individualized Instruction in STEM Learning Environments (1726550) and CAREER: Improving Adaptive Decision Making in Interactive Learning Environments (1651909).

\bibliographystyle{apacite}

\setlength{\bibleftmargin}{.125in}
\setlength{\bibindent}{-\bibleftmargin}

\bibliography{cogsci2021}

\begin{thebibliography}{}

\bibitem [\protect \citeauthoryear {%
Abdelshiheed%
\ \protect \BOthers {.}}{%
Abdelshiheed%
\ \protect \BOthers {.}}{%
{\protect \APACyear {2020}}%
}]{%
abdelshiheed2020metacognition}
\APACinsertmetastar {%
abdelshiheed2020metacognition}%
\begin{APACrefauthors}%
Abdelshiheed, M.%
\BCBT {}\ \BOthersPeriod {.}
\end{APACrefauthors}%
\unskip\
\newblock
\APACrefYearMonthDay{2020}{}{}.
\newblock
{\BBOQ}\APACrefatitle {Metacognition and Motivation: The Role of Time-Awareness
  in Preparation for Future Learning} {Metacognition and motivation: The role
  of time-awareness in preparation for future learning}.{\BBCQ}
\newblock
\BIn{} \APACrefbtitle {Proceedings of the 42nd annual conference of the
  cognitive science society.} {Proceedings of the 42nd annual conference of the
  cognitive science society.}
\PrintBackRefs{\CurrentBib}

\bibitem [\protect \citeauthoryear {%
Alexander%
\ \protect \BOthers {.}}{%
Alexander%
\ \protect \BOthers {.}}{%
{\protect \APACyear {1998}}%
}]{%
alexander1998perspectiveSWITCH}
\APACinsertmetastar {%
alexander1998perspectiveSWITCH}%
\begin{APACrefauthors}%
Alexander, P\BPBI A.%
\BCBT {}\ \BOthersPeriod {.}
\end{APACrefauthors}%
\unskip\
\newblock
\APACrefYearMonthDay{1998}{}{}.
\newblock
{\BBOQ}\APACrefatitle {A perspective on strategy research: Progress and
  prospects} {A perspective on strategy research: Progress and
  prospects}.{\BBCQ}
\newblock
\APACjournalVolNumPages{Educ. Psychol. Rev.}{10}{2}{129--154}.
\PrintBackRefs{\CurrentBib}

\bibitem [\protect \citeauthoryear {%
Ardiale%
\ \BBA {} Lemaire%
}{%
Ardiale%
\ \BBA {} Lemaire%
}{%
{\protect \APACyear {2012}}%
}]{%
ardiale2012withinSWITCH}
\APACinsertmetastar {%
ardiale2012withinSWITCH}%
\begin{APACrefauthors}%
Ardiale, E.%
\BCBT {}\ \BBA {} Lemaire, P.%
\end{APACrefauthors}%
\unskip\
\newblock
\APACrefYearMonthDay{2012}{}{}.
\newblock
{\BBOQ}\APACrefatitle {Within-item strategy switching: An age comparative study
  in adults.} {Within-item strategy switching: An age comparative study in
  adults.}{\BBCQ}
\newblock
\APACjournalVolNumPages{Psychology and aging}{27}{4}{1138}.
\PrintBackRefs{\CurrentBib}

\bibitem [\protect \citeauthoryear {%
Belenky%
\ \BBA {} Nokes-Malach%
}{%
Belenky%
\ \BBA {} Nokes-Malach%
}{%
{\protect \APACyear {2012}}%
}]{%
belenky2012motivationAndTransferPFL}
\APACinsertmetastar {%
belenky2012motivationAndTransferPFL}%
\begin{APACrefauthors}%
Belenky, D.%
\BCBT {}\ \BBA {} Nokes-Malach, T\BPBI J.%
\end{APACrefauthors}%
\unskip\
\newblock
\APACrefYearMonthDay{2012}{}{}.
\newblock
{\BBOQ}\APACrefatitle {Motivation and transfer: The role of mastery-approach
  goals in preparation for future learning} {Motivation and transfer: The role
  of mastery-approach goals in preparation for future learning}.{\BBCQ}
\newblock
\APACjournalVolNumPages{J. Learn. Sci.}{21}{3}{399--432}.
\PrintBackRefs{\CurrentBib}

\bibitem [\protect \citeauthoryear {%
Bransford%
\ \BBA {} Schwartz%
}{%
Bransford%
\ \BBA {} Schwartz%
}{%
{\protect \APACyear {1999}}%
}]{%
bransford1999transferRethinking}
\APACinsertmetastar {%
bransford1999transferRethinking}%
\begin{APACrefauthors}%
Bransford, J\BPBI D.%
\BCBT {}\ \BBA {} Schwartz, D\BPBI L.%
\end{APACrefauthors}%
\unskip\
\newblock
\APACrefYearMonthDay{1999}{}{}.
\newblock
{\BBOQ}\APACrefatitle {Rethinking transfer: A simple proposal with multiple
  implications} {Rethinking transfer: A simple proposal with multiple
  implications}.{\BBCQ}
\newblock
\APACjournalVolNumPages{Review of research in education}{24}{1}{61--100}.
\PrintBackRefs{\CurrentBib}

\bibitem [\protect \citeauthoryear {%
Cardelle-Elawar%
}{%
Cardelle-Elawar%
}{%
{\protect \APACyear {1992}}%
}]{%
cardelle1992effectsMETACOGNITIVE}
\APACinsertmetastar {%
cardelle1992effectsMETACOGNITIVE}%
\begin{APACrefauthors}%
Cardelle-Elawar, M.%
\end{APACrefauthors}%
\unskip\
\newblock
\APACrefYearMonthDay{1992}{}{}.
\newblock
{\BBOQ}\APACrefatitle {Effects of teaching metacognitive skills to students
  with low mathematics ability} {Effects of teaching metacognitive skills to
  students with low mathematics ability}.{\BBCQ}
\newblock
\APACjournalVolNumPages{Teaching and teacher education}{8}{2}{109--121}.
\PrintBackRefs{\CurrentBib}

\bibitem [\protect \citeauthoryear {%
Chambres%
\ \protect \BOthers {.}}{%
Chambres%
\ \protect \BOthers {.}}{%
{\protect \APACyear {2002}}%
}]{%
chambres2002metacognitionStrategySelection}
\APACinsertmetastar {%
chambres2002metacognitionStrategySelection}%
\begin{APACrefauthors}%
Chambres, P.%
\BCBT {}\ \BOthersPeriod {.}
\end{APACrefauthors}%
\unskip\
\newblock
\APACrefYear{2002}.
\newblock
\APACrefbtitle {Metacognition: Process, function, and use} {Metacognition:
  Process, function, and use}.
\newblock
\APACaddressPublisher{}{Kluwer Academic Publishers}.
\PrintBackRefs{\CurrentBib}

\bibitem [\protect \citeauthoryear {%
Chi%
\ \BBA {} VanLehn%
}{%
Chi%
\ \BBA {} VanLehn%
}{%
{\protect \APACyear {2010}}%
}]{%
chi2010backward2metacogStrategyINSTRUCTION}
\APACinsertmetastar {%
chi2010backward2metacogStrategyINSTRUCTION}%
\begin{APACrefauthors}%
Chi, M.%
\BCBT {}\ \BBA {} VanLehn, K.%
\end{APACrefauthors}%
\unskip\
\newblock
\APACrefYearMonthDay{2010}{}{}.
\newblock
{\BBOQ}\APACrefatitle {Meta-Cognitive Strategy Instruction in Intelligent
  Tutoring Systems: How, When, and Why.} {Meta-cognitive strategy instruction
  in intelligent tutoring systems: How, when, and why.}{\BBCQ}
\newblock
\APACjournalVolNumPages{Educational Technology \& Society}{13}{1}{25--39}.
\PrintBackRefs{\CurrentBib}

\bibitem [\protect \citeauthoryear {%
Cronbach%
\ \BBA {} Snow%
}{%
Cronbach%
\ \BBA {} Snow%
}{%
{\protect \APACyear {1977}}%
}]{%
cronbach1977aptitudes}
\APACinsertmetastar {%
cronbach1977aptitudes}%
\begin{APACrefauthors}%
Cronbach, L\BPBI J.%
\BCBT {}\ \BBA {} Snow, R\BPBI E.%
\end{APACrefauthors}%
\unskip\
\newblock
\APACrefYear{1977}.
\newblock
\APACrefbtitle {Aptitudes and instructional methods: A handbook for research on
  interactions.} {Aptitudes and instructional methods: A handbook for research
  on interactions.}
\PrintBackRefs{\CurrentBib}

\bibitem [\protect \citeauthoryear {%
de Boer%
\ \protect \BOthers {.}}{%
de Boer%
\ \protect \BOthers {.}}{%
{\protect \APACyear {2018}}%
}]{%
de2018longSWITCH-INSTRUCTION}
\APACinsertmetastar {%
de2018longSWITCH-INSTRUCTION}%
\begin{APACrefauthors}%
de Boer, H.%
\BCBT {}\ \BOthersPeriod {.}
\end{APACrefauthors}%
\unskip\
\newblock
\APACrefYearMonthDay{2018}{}{}.
\newblock
{\BBOQ}\APACrefatitle {Long-term effects of metacognitive strategy instruction
  on student academic performance: A meta-analysis} {Long-term effects of
  metacognitive strategy instruction on student academic performance: A
  meta-analysis}.{\BBCQ}
\newblock
\APACjournalVolNumPages{Educational Research Review}{24}{}{98--115}.
\PrintBackRefs{\CurrentBib}

\bibitem [\protect \citeauthoryear {%
Gerjets%
\ \protect \BOthers {.}}{%
Gerjets%
\ \protect \BOthers {.}}{%
{\protect \APACyear {2004}}%
}]{%
gerjets2004designingINSTRUCTION}
\APACinsertmetastar {%
gerjets2004designingINSTRUCTION}%
\begin{APACrefauthors}%
Gerjets, P.%
\BCBT {}\ \BOthersPeriod {.}
\end{APACrefauthors}%
\unskip\
\newblock
\APACrefYearMonthDay{2004}{}{}.
\newblock
{\BBOQ}\APACrefatitle {Designing instructional examples to reduce intrinsic
  cognitive load: Molar versus modular presentation of solution procedures}
  {Designing instructional examples to reduce intrinsic cognitive load: Molar
  versus modular presentation of solution procedures}.{\BBCQ}
\newblock
\APACjournalVolNumPages{Instr. Sci.}{32}{}{33--58}.
\PrintBackRefs{\CurrentBib}

\bibitem [\protect \citeauthoryear {%
Glogger-Frey%
\ \protect \BOthers {.}}{%
Glogger-Frey%
\ \protect \BOthers {.}}{%
{\protect \APACyear {2015}}%
}]{%
glogger2015INSTRUCTION}
\APACinsertmetastar {%
glogger2015INSTRUCTION}%
\begin{APACrefauthors}%
Glogger-Frey, I.%
\BCBT {}\ \BOthersPeriod {.}
\end{APACrefauthors}%
\unskip\
\newblock
\APACrefYearMonthDay{2015}{}{}.
\newblock
{\BBOQ}\APACrefatitle {Inventing a solution and studying a worked solution
  prepare differently for learning from direct instruction} {Inventing a
  solution and studying a worked solution prepare differently for learning from
  direct instruction}.{\BBCQ}
\newblock
\APACjournalVolNumPages{Learning and Instruction}{39}{}{72--87}.
\PrintBackRefs{\CurrentBib}

\bibitem [\protect \citeauthoryear {%
Hartman%
}{%
Hartman%
}{%
{\protect \APACyear {2001}}%
}]{%
hartman2001developingSWITCH}
\APACinsertmetastar {%
hartman2001developingSWITCH}%
\begin{APACrefauthors}%
Hartman, H\BPBI J.%
\end{APACrefauthors}%
\unskip\
\newblock
\APACrefYearMonthDay{2001}{}{}.
\newblock
{\BBOQ}\APACrefatitle {Developing students’ metacognitive knowledge and
  skills} {Developing students’ metacognitive knowledge and skills}.{\BBCQ}
\newblock
\APACjournalVolNumPages{Metacognition in learning and instruction}{}{}{33--68}.
\PrintBackRefs{\CurrentBib}

\bibitem [\protect \citeauthoryear {%
Ju%
, Zhou%
, Abdelshiheed%
, Barnes%
\BCBL {}\ \BBA {} Chi%
}{%
Ju%
\ \protect \BOthers {.}}{%
{\protect \APACyear {2021}}%
}]{%
song2021AIED}
\APACinsertmetastar {%
song2021AIED}%
\begin{APACrefauthors}%
Ju, S.%
, Zhou, G.%
, Abdelshiheed, M.%
, Barnes, T.%
\BCBL {}\ \BBA {} Chi, M.%
\end{APACrefauthors}%
\unskip\
\newblock
\APACrefYearMonthDay{2021}{}{}.
\newblock
{\BBOQ}\APACrefatitle {Evaluating Critical Reinforcement Learning Framework In
  the Field} {Evaluating critical reinforcement learning framework in the
  field}.{\BBCQ}
\newblock
\BIn{} \APACrefbtitle {International Conference on Artificial Intelligence in
  Education.} {International conference on artificial intelligence in
  education.}
\PrintBackRefs{\CurrentBib}

\bibitem [\protect \citeauthoryear {%
Larkin%
\ \protect \BOthers {.}}{%
Larkin%
\ \protect \BOthers {.}}{%
{\protect \APACyear {1980}}%
}]{%
larkin1980expertINSTRUCTION}
\APACinsertmetastar {%
larkin1980expertINSTRUCTION}%
\begin{APACrefauthors}%
Larkin, J.%
\BCBT {}\ \BOthersPeriod {.}
\end{APACrefauthors}%
\unskip\
\newblock
\APACrefYearMonthDay{1980}{}{}.
\newblock
{\BBOQ}\APACrefatitle {Expert and novice performance in solving physics
  problems} {Expert and novice performance in solving physics problems}.{\BBCQ}
\newblock
\APACjournalVolNumPages{Science}{208}{}{1335--1342}.
\PrintBackRefs{\CurrentBib}

\bibitem [\protect \citeauthoryear {%
Likourezos%
\ \BBA {} Kalyuga%
}{%
Likourezos%
\ \BBA {} Kalyuga%
}{%
{\protect \APACyear {2017}}%
}]{%
likourezos2017instruction}
\APACinsertmetastar {%
likourezos2017instruction}%
\begin{APACrefauthors}%
Likourezos, V.%
\BCBT {}\ \BBA {} Kalyuga, S.%
\end{APACrefauthors}%
\unskip\
\newblock
\APACrefYearMonthDay{2017}{}{}.
\newblock
{\BBOQ}\APACrefatitle {Instruction-first and problem-solving-first approaches:
  alternative pathways to learning complex tasks} {Instruction-first and
  problem-solving-first approaches: alternative pathways to learning complex
  tasks}.{\BBCQ}
\newblock
\APACjournalVolNumPages{Instr. Sci.}{45}{}{195--219}.
\PrintBackRefs{\CurrentBib}

\bibitem [\protect \citeauthoryear {%
Mylopoulos%
\ \protect \BOthers {.}}{%
Mylopoulos%
\ \protect \BOthers {.}}{%
{\protect \APACyear {2016}}%
}]{%
mylopoulos2016PFL}
\APACinsertmetastar {%
mylopoulos2016PFL}%
\begin{APACrefauthors}%
Mylopoulos, M.%
\BCBT {}\ \BOthersPeriod {.}
\end{APACrefauthors}%
\unskip\
\newblock
\APACrefYearMonthDay{2016}{}{}.
\newblock
{\BBOQ}\APACrefatitle {Preparation for future learning: a missing competency in
  health professions education?} {Preparation for future learning: a missing
  competency in health professions education?}{\BBCQ}
\newblock
\APACjournalVolNumPages{Medical education}{50}{1}{115--123}.
\PrintBackRefs{\CurrentBib}

\bibitem [\protect \citeauthoryear {%
Paas%
\ \protect \BOthers {.}}{%
Paas%
\ \protect \BOthers {.}}{%
{\protect \APACyear {1994}}%
}]{%
paas1994variabilityINSTRUCTION}
\APACinsertmetastar {%
paas1994variabilityINSTRUCTION}%
\begin{APACrefauthors}%
Paas, F\BPBI G.%
\BCBT {}\ \BOthersPeriod {.}
\end{APACrefauthors}%
\unskip\
\newblock
\APACrefYearMonthDay{1994}{}{}.
\newblock
{\BBOQ}\APACrefatitle {Variability of worked examples and transfer of
  geometrical problem-solving skills: A cognitive-load approach.} {Variability
  of worked examples and transfer of geometrical problem-solving skills: A
  cognitive-load approach.}{\BBCQ}
\newblock
\APACjournalVolNumPages{J. Educ. Psychol.}{86}{1}{122}.
\PrintBackRefs{\CurrentBib}

\bibitem [\protect \citeauthoryear {%
Pintrich%
}{%
Pintrich%
}{%
{\protect \APACyear {2002}}%
}]{%
pintrich2002roleSWITCH}
\APACinsertmetastar {%
pintrich2002roleSWITCH}%
\begin{APACrefauthors}%
Pintrich, P\BPBI R.%
\end{APACrefauthors}%
\unskip\
\newblock
\APACrefYearMonthDay{2002}{}{}.
\newblock
{\BBOQ}\APACrefatitle {The role of metacognitive knowledge in learning,
  teaching, and assessing} {The role of metacognitive knowledge in learning,
  teaching, and assessing}.{\BBCQ}
\newblock
\APACjournalVolNumPages{Theory into practice}{41}{4}{219--225}.
\PrintBackRefs{\CurrentBib}

\bibitem [\protect \citeauthoryear {%
Priest%
\ \BBA {} Lindsay%
}{%
Priest%
\ \BBA {} Lindsay%
}{%
{\protect \APACyear {1992}}%
}]{%
priest1992newINSTRUCTION}
\APACinsertmetastar {%
priest1992newINSTRUCTION}%
\begin{APACrefauthors}%
Priest, A.%
\BCBT {}\ \BBA {} Lindsay, R.%
\end{APACrefauthors}%
\unskip\
\newblock
\APACrefYearMonthDay{1992}{}{}.
\newblock
{\BBOQ}\APACrefatitle {New light on novice—expert differences in physics
  problem solving} {New light on novice—expert differences in physics problem
  solving}.{\BBCQ}
\newblock
\APACjournalVolNumPages{British journal of Psychology}{83}{3}{389--405}.
\PrintBackRefs{\CurrentBib}

\bibitem [\protect \citeauthoryear {%
Renkl%
\ \BBA {} Atkinson%
}{%
Renkl%
\ \BBA {} Atkinson%
}{%
{\protect \APACyear {2010}}%
}]{%
renkl2010learningINSTRUCTION}
\APACinsertmetastar {%
renkl2010learningINSTRUCTION}%
\begin{APACrefauthors}%
Renkl, A.%
\BCBT {}\ \BBA {} Atkinson, R\BPBI K.%
\end{APACrefauthors}%
\unskip\
\newblock
\APACrefYearMonthDay{2010}{}{}.
\newblock
{\BBOQ}\APACrefatitle {Learning from worked-out examples and problem solving.}
  {Learning from worked-out examples and problem solving.}{\BBCQ}
\newblock

\PrintBackRefs{\CurrentBib}

\bibitem [\protect \citeauthoryear {%
Roberts%
\ \BBA {} Erdos%
}{%
Roberts%
\ \BBA {} Erdos%
}{%
{\protect \APACyear {1993}}%
}]{%
roberts1993metacogDefinitionStrategySelection2}
\APACinsertmetastar {%
roberts1993metacogDefinitionStrategySelection2}%
\begin{APACrefauthors}%
Roberts, M\BPBI J.%
\BCBT {}\ \BBA {} Erdos, G.%
\end{APACrefauthors}%
\unskip\
\newblock
\APACrefYearMonthDay{1993}{}{}.
\newblock
{\BBOQ}\APACrefatitle {Strategy selection and metacognition} {Strategy
  selection and metacognition}.{\BBCQ}
\newblock
\APACjournalVolNumPages{Educational Psychology}{13}{}{259--266}.
\PrintBackRefs{\CurrentBib}

\bibitem [\protect \citeauthoryear {%
Russell%
\ \BBA {} Norvig%
}{%
Russell%
\ \BBA {} Norvig%
}{%
{\protect \APACyear {2010}}%
}]{%
russell2010artificial}
\APACinsertmetastar {%
russell2010artificial}%
\begin{APACrefauthors}%
Russell, S\BPBI J.%
\BCBT {}\ \BBA {} Norvig, P.%
\end{APACrefauthors}%
\unskip\
\newblock
\APACrefYearMonthDay{2010}{}{}.
\newblock
{\BBOQ}\APACrefatitle {Artificial intelligence: a modern approach} {Artificial
  intelligence: a modern approach}.{\BBCQ}
\newblock

\PrintBackRefs{\CurrentBib}

\bibitem [\protect \citeauthoryear {%
Sanz~Ausin%
\ \protect \BOthers {.}}{%
Sanz~Ausin%
\ \protect \BOthers {.}}{%
{\protect \APACyear {2019}}%
}]{%
markelATI}
\APACinsertmetastar {%
markelATI}%
\begin{APACrefauthors}%
Sanz~Ausin, M.%
\BCBT {}\ \BOthersPeriod {.}
\end{APACrefauthors}%
\unskip\
\newblock
\APACrefYearMonthDay{2019}{}{}.
\newblock
{\BBOQ}\APACrefatitle {Leveraging Deep Reinforcement Learning for Pedagogical
  Policy Induction in an Intelligent Tutoring System} {Leveraging deep
  reinforcement learning for pedagogical policy induction in an intelligent
  tutoring system}.{\BBCQ}
\newblock
\BIn{} \APACrefbtitle {Proceedings of the 12th International Conference on
  Educational Data Mining.} {Proceedings of the 12th international conference
  on educational data mining.}
\PrintBackRefs{\CurrentBib}

\bibitem [\protect \citeauthoryear {%
Schwonke%
\ \protect \BOthers {.}}{%
Schwonke%
\ \protect \BOthers {.}}{%
{\protect \APACyear {2009}}%
}]{%
schwonke2009WE}
\APACinsertmetastar {%
schwonke2009WE}%
\begin{APACrefauthors}%
Schwonke, R.%
\BCBT {}\ \BOthersPeriod {.}
\end{APACrefauthors}%
\unskip\
\newblock
\APACrefYearMonthDay{2009}{}{}.
\newblock
{\BBOQ}\APACrefatitle {The worked-example effect: Not an artefact of lousy
  control conditions} {The worked-example effect: Not an artefact of lousy
  control conditions}.{\BBCQ}
\newblock
\APACjournalVolNumPages{Computers in human behavior}{25}{2}{258--266}.
\PrintBackRefs{\CurrentBib}

\bibitem [\protect \citeauthoryear {%
Shrestha%
\ \protect \BOthers {.}}{%
Shrestha%
\ \protect \BOthers {.}}{%
{\protect \APACyear {2013}}%
}]{%
shrestha2013usingINSTRUCTION}
\APACinsertmetastar {%
shrestha2013usingINSTRUCTION}%
\begin{APACrefauthors}%
Shrestha, A.%
\BCBT {}\ \BOthersPeriod {.}
\end{APACrefauthors}%
\unskip\
\newblock
\APACrefYearMonthDay{2013}{}{}.
\newblock
{\BBOQ}\APACrefatitle {Using point-of-view video modeling and forward chaining
  to teach a functional self-help skill to a child with autism} {Using
  point-of-view video modeling and forward chaining to teach a functional
  self-help skill to a child with autism}.{\BBCQ}
\newblock
\APACjournalVolNumPages{J. Behav. Educ.}{22}{}{157--167}.
\PrintBackRefs{\CurrentBib}

\bibitem [\protect \citeauthoryear {%
Simon%
\ \BBA {} Simon%
}{%
Simon%
\ \BBA {} Simon%
}{%
{\protect \APACyear {1978}}%
}]{%
simon1978individualINSTRUCTION}
\APACinsertmetastar {%
simon1978individualINSTRUCTION}%
\begin{APACrefauthors}%
Simon, D\BPBI P.%
\BCBT {}\ \BBA {} Simon, H\BPBI A.%
\end{APACrefauthors}%
\unskip\
\newblock
\APACrefYearMonthDay{1978}{}{}.
\newblock
{\BBOQ}\APACrefatitle {Individual differences in solving physics problems.}
  {Individual differences in solving physics problems.}{\BBCQ}
\newblock

\PrintBackRefs{\CurrentBib}

\bibitem [\protect \citeauthoryear {%
Sp{\"o}rer%
\ \protect \BOthers {.}}{%
Sp{\"o}rer%
\ \protect \BOthers {.}}{%
{\protect \APACyear {2009}}%
}]{%
sporer2009improvingINSTRUCTION}
\APACinsertmetastar {%
sporer2009improvingINSTRUCTION}%
\begin{APACrefauthors}%
Sp{\"o}rer, N.%
\BCBT {}\ \BOthersPeriod {.}
\end{APACrefauthors}%
\unskip\
\newblock
\APACrefYearMonthDay{2009}{}{}.
\newblock
{\BBOQ}\APACrefatitle {Improving students' reading comprehension skills:
  Effects of strategy instruction and reciprocal teaching} {Improving students'
  reading comprehension skills: Effects of strategy instruction and reciprocal
  teaching}.{\BBCQ}
\newblock
\APACjournalVolNumPages{Learning and instruction}{19}{3}{272--286}.
\PrintBackRefs{\CurrentBib}

\bibitem [\protect \citeauthoryear {%
Sweller%
\ \BBA {} Cooper%
}{%
Sweller%
\ \BBA {} Cooper%
}{%
{\protect \APACyear {1985}}%
}]{%
sweller1985WE}
\APACinsertmetastar {%
sweller1985WE}%
\begin{APACrefauthors}%
Sweller, J.%
\BCBT {}\ \BBA {} Cooper, G\BPBI A.%
\end{APACrefauthors}%
\unskip\
\newblock
\APACrefYearMonthDay{1985}{}{}.
\newblock
{\BBOQ}\APACrefatitle {The use of worked examples as a substitute for problem
  solving in learning algebra} {The use of worked examples as a substitute for
  problem solving in learning algebra}.{\BBCQ}
\newblock
\APACjournalVolNumPages{Cognition and instruction}{2}{1}{59--89}.
\PrintBackRefs{\CurrentBib}

\bibitem [\protect \citeauthoryear {%
Taillan%
\ \protect \BOthers {.}}{%
Taillan%
\ \protect \BOthers {.}}{%
{\protect \APACyear {2015}}%
}]{%
taillan2015relationshipsSWITCH}
\APACinsertmetastar {%
taillan2015relationshipsSWITCH}%
\begin{APACrefauthors}%
Taillan, J.%
\BCBT {}\ \BOthersPeriod {.}
\end{APACrefauthors}%
\unskip\
\newblock
\APACrefYearMonthDay{2015}{}{}.
\newblock
{\BBOQ}\APACrefatitle {Relationships between strategy switching and strategy
  switch costs in young and older adults: A study in arithmetic problem
  solving} {Relationships between strategy switching and strategy switch costs
  in young and older adults: A study in arithmetic problem solving}.{\BBCQ}
\newblock
\APACjournalVolNumPages{Experimental aging research}{41}{2}{136--156}.
\PrintBackRefs{\CurrentBib}

\bibitem [\protect \citeauthoryear {%
Van~Gog%
\ \protect \BOthers {.}}{%
Van~Gog%
\ \protect \BOthers {.}}{%
{\protect \APACyear {2004}}%
}]{%
van2004processINSTRUCTION}
\APACinsertmetastar {%
van2004processINSTRUCTION}%
\begin{APACrefauthors}%
Van~Gog, T.%
\BCBT {}\ \BOthersPeriod {.}
\end{APACrefauthors}%
\unskip\
\newblock
\APACrefYearMonthDay{2004}{}{}.
\newblock
{\BBOQ}\APACrefatitle {Process-oriented worked examples: Improving transfer
  performance through enhanced understanding} {Process-oriented worked
  examples: Improving transfer performance through enhanced
  understanding}.{\BBCQ}
\newblock
\APACjournalVolNumPages{Instr. Sci.}{32}{1}{83--98}.
\PrintBackRefs{\CurrentBib}

\bibitem [\protect \citeauthoryear {%
Veenman%
\ \protect \BOthers {.}}{%
Veenman%
\ \protect \BOthers {.}}{%
{\protect \APACyear {2004}}%
}]{%
veenman2004relationMetacognitivePFL}
\APACinsertmetastar {%
veenman2004relationMetacognitivePFL}%
\begin{APACrefauthors}%
Veenman, M\BPBI V.%
\BCBT {}\ \BOthersPeriod {.}
\end{APACrefauthors}%
\unskip\
\newblock
\APACrefYearMonthDay{2004}{}{}.
\newblock
{\BBOQ}\APACrefatitle {The relation between intellectual and metacognitive
  skills from a developmental perspective} {The relation between intellectual
  and metacognitive skills from a developmental perspective}.{\BBCQ}
\newblock
\APACjournalVolNumPages{Learning and instruction}{14}{1}{89--109}.
\PrintBackRefs{\CurrentBib}

\bibitem [\protect \citeauthoryear {%
Winne%
\ \BBA {} Azevedo%
}{%
Winne%
\ \BBA {} Azevedo%
}{%
{\protect \APACyear {2014}}%
}]{%
winne2014switchMetacognitiveSWITCH}
\APACinsertmetastar {%
winne2014switchMetacognitiveSWITCH}%
\begin{APACrefauthors}%
Winne, P\BPBI H.%
\BCBT {}\ \BBA {} Azevedo, R.%
\end{APACrefauthors}%
\unskip\
\newblock
\APACrefYearMonthDay{2014}{}{}.
\newblock
{\BBOQ}\APACrefatitle {Metacognition} {Metacognition}.{\BBCQ}
\newblock
\BIn{} R\BPBI K.~Sawyer\ (\BED), \APACrefbtitle {The Cambridge Handbook of the
  Learning Sciences} {The cambridge handbook of the learning sciences}\
  (\PrintOrdinal{2}\ \BEd).
\newblock
\APACaddressPublisher{}{Cambridge University Press}.
\PrintBackRefs{\CurrentBib}

\bibitem [\protect \citeauthoryear {%
Zepeda%
\ \protect \BOthers {.}}{%
Zepeda%
\ \protect \BOthers {.}}{%
{\protect \APACyear {2015}}%
}]{%
zepeda2015INSTRUCTION}
\APACinsertmetastar {%
zepeda2015INSTRUCTION}%
\begin{APACrefauthors}%
Zepeda, C\BPBI D.%
\BCBT {}\ \BOthersPeriod {.}
\end{APACrefauthors}%
\unskip\
\newblock
\APACrefYearMonthDay{2015}{}{}.
\newblock
{\BBOQ}\APACrefatitle {Direct instruction of metacognition benefits adolescent
  science learning, transfer, and motivation: An in vivo study.} {Direct
  instruction of metacognition benefits adolescent science learning, transfer,
  and motivation: An in vivo study.}{\BBCQ}
\newblock
\APACjournalVolNumPages{J. Educ. Psychol.}{107}{4}{954}.
\PrintBackRefs{\CurrentBib}

\bibitem [\protect \citeauthoryear {%
Zhou%
\ \BBA {} Chi%
}{%
Zhou%
\ \BBA {} Chi%
}{%
{\protect \APACyear {2017}}%
}]{%
zhou2017impact}
\APACinsertmetastar {%
zhou2017impact}%
\begin{APACrefauthors}%
Zhou, G.%
\BCBT {}\ \BBA {} Chi, M.%
\end{APACrefauthors}%
\unskip\
\newblock
\APACrefYearMonthDay{2017}{}{}.
\newblock
{\BBOQ}\APACrefatitle {The Impact of Decision Agency \& Granularity on Aptitude
  Treatment Interaction in Tutoring.} {The impact of decision agency \&
  granularity on aptitude treatment interaction in tutoring.}{\BBCQ}
\newblock
\BIn{} \APACrefbtitle {CogSci.} {Cogsci.}
\PrintBackRefs{\CurrentBib}

\end{thebibliography}

\end{document}